# A Chronology of Torah Cryptography

Grenville J. Croll
grenville@spreadsheetrisks.com
v1.10

**Regarding some papers and notes submitted to, or presented at, the second congress of the International Torah Codes Society in Jerusalem, Israel, June 2000.**

## 1 INTRODUCTION

Sometime during the early part of 1997 the author became aware of the controversial paper "Equidistant Letter Sequences in the Book of Genesis" [Witztum, Rips and Rosenberg, 1994] (WRR). This paper brought to the attention of a wider audience the possibility that the Torah (the first five books of the Hebrew Bible) was some kind of cryptographic material. Though this work was countered by a later and very comprehensive rebuttal [McKay, 1998] [McKay et al, 1999], the search for a deeper scientific understanding of the cryptographic nature of the Torah continues [Haralick, 2006a, 2006b].

Whilst the public debate about WRR's work progressed in the late 1990's, I quietly pursued my own, very different, ideas about the problem of how to possibly decode the Torah. Based upon some simple mathematical and geometrical principles, I outlined my *a priori* ideas in a short book, entitled "The Keys and the Seven Seals" [Croll, 1997], written during the summer of 1997. The book outlined a method of potentially re-arranging the characters of the existing Torah into a completely new lucid document, a New Torah, using a simple algorithm. Thirty copies of the book were printed, bound and distributed to a variety of people, including WRR.

This book was converted into a short mathematical paper and submitted to the Computer Journal of the British Computer Society. It was rejected. However, I am grateful to the editor at the time, Professor C. van Rijsbergen, for putting the paper through peer review and to the referee for his or her constructive comments.

Having written my book, it became very clear that even to attempt to decode the holiest book of the Jewish tradition was a very serious matter. The thought of possibly succeeding in the way that Alan Turing [Hodges, 1992] did with his decryption of the German Enigma codes during World War II did not bear contemplation.

After a pause for contemplation of about one year, I started work on my own decode project, and obtained a digital copy of the Torah for the first time [Computronic, 1997]. My work initially focused on testing certain permutations of the Torah for randomness [Croll, 2000a], using my own and other people's tests of disorder [Pincus, 1991]. This first paper on randomness was circulated widely amongst the community of people involved in the Torah Codes issue. It was submitted to the Journal of the Royal Statistical Society and rejected. I express my gratitude to Dr Randall Ingermanson [Ingermanson, 1999], for identifying & clearly elucidating the fault in that work, and to Professor Brendan McKay for his suggestions regarding the circumvention of that fault. Unfortunately I have not yet had time to revisit that particular project, mainly because I am fairly sure that the results would not now be particularly interesting. I thank Dr. Steve Pincus, developer of the Approximate Entropy function for his encouragement and copies of his most interesting papers.

In April 1999, I turned my attention to the small piece of text (T2) located between the inverted letter nuns in Numbers 10 verses 34-35. This work was extremely fruitful. I was able to show how the *a priori* algorithm outlined in my book could be used to permute this small section of text into multiple alternate readable forms. By securing the services of a reputable commercial translation agency in London, it is clear that these and similar permutations are readable to a greater or lesser extent. The initial decryption strategy was to use the parameters used to generate any lucid T2 sequences as parameters for a plain text decryption of the whole Torah. Unfortunately, the project became compute bound due to combinatorial complexity, and constrained by time and the lack of any available Hebrew speakers & readers. Despite the lack of later success on the whole Torah, the cryptanalysis of T2 was successful as we showed that it was possible, at least in principle, to algorithmically permute a small part of the Torah ("A Sefer" or book) into one or more readable documents using a simple algorithm..

In the latter part of 1999 I received, from Prof. Eliyahu Rips, a call for papers for the second congress of the International Torah Codes Society which was to be held in Jerusalem in June 2000. I sent three full papers [Croll, 2000a,b,c] and two short notes (essentially Sections 2 & 3 of this paper) to Dr Moshe Katz, the congress organiser, for consideration. I presented two full papers [Croll, 2000b,c] on the 5th & 6th June 2000 respectively. The first short note was discussed briefly in a conversation with Prof. Robert Haralick and Dr Harold Gans. Prof. Haralick intimated that he had checked my 1839 observation "from both ends" by writing his own computer program and by visiting the web site of the National Physical Laboratory to check the number with regard to the known physical constants. The second short note was neither discussed nor mentioned at the conference, though there was a suggestion that Prof. Eliyahu Rips was now searching for compound sequences of a similar length.

On the second day of the conference, I was surprised to have been shown a large scale and very detailed plan of the Temple Mount in Jerusalem. There was some discussion regarding the precise geometrical layout of the site. I was also shown a large scale map of the Eastern Mediterranean region, with the Biblical lands of Israel clearly colour coded according to each tribe. The relevance of these plans in a cryptographic conference was not clear at the time. However, a recent footnote regarding the Torah in Professor T. W. Körner's taught course on Codes and Cryptography [Körner, 2011], part of the Cambridge University Mathematics Tripos, did shed some further light on the relevance of my being shown these plans.

Following the ITCS 2000 congress, there was no response to my request for information regarding the publication of the conference proceedings. In early 2001, I submitted the two full papers for consideration regarding publication in the academic journal Cryptologia [Nichols, 1998]. After a lengthy delay the papers were rejected as the journal editor, Professor Brian Winkel, could not find referees interested in reviewing the work.

Following some research on the internet, I became aware of the existence of the Alternative Natural Philosophy Society (ANPA). As a result of an enquiry on an internet discussion group, Professor Brian Josephson kindly advised me by email in August 2004 that the 25th annual ANPA meeting was currently in progress at Wesley College, Jesus Lane, Cambridge. I attended the latter part of the 25th meeting where I was able to enjoy a couple of presentations by delegates, and a short speech by Dr Pierre Noyes, founder of the association, celebrating its first 25 years.

The following year, I was able to present my approach to the Torah Decryption problem at the 26th ANPA meeting which was attended by a small audience which included Prof. Louis Kauffman, the late Prof. Clive Kilmister and Prof. John H. Conway (who commented upon the ineluctability of my work). Following my own presentation, I was privileged to listen to Prof. Conway's presentation on the Free Will Theorem [Conway & Kochen, 2006]. My two full papers were subsequently published in the ANPA conference proceedings and uploaded to Arxiv in 2007.

The purpose of this paper is therefore to document the main texts and sequences of events so far regarding my work on the Torah Decryption problem. Section 2 outlines some coincidences which suggest that there may well be an element of design in the Torah due to the disclosure of one or possibly two key constants relevant within the domain of theoretical physics. Section 3 documents a further set of coincidences that also suggest that there may well be an element of design in the Torah. In section 4 we summarise our work and in section 5 outline some ideas for future work. We include in the Appendices the bit map for a message which was transmitted from Arecibo in 1974, the plaque on the Pioneer 10 & 11 spacecraft and the transliteration table between the Hebrew and English alphabets and symbols used in this paper.

## 2 SOME 'COINCIDENCES' IN THE TORAH

In our randomness paper [Croll, 2000a], we investigate how disordered various documents become when they are permuted using our Algorithm One. Documents investigated include the Torah, other parts of the Hebrew Bible and the Bible in English. The method used for testing randomness on these documents is also tested against some well known irrational and transcendental numbers including pi and e, which have a range of well researched properties in this area [Pincus & Kalman, 1997].

Amongst a number of results, we show that these simple permutations of the Torah are completely random and that statistically speaking, there is no evidence that the Torah, when permuted by Algorithm One, is in any way different from an artificially created random jumble of Hebrew letters.

The assessment of randomness in our paper involves the scanning of a document to reveal what is the most frequently occurring set of two, three or four letters. A simple set of functions called MaxPair, MaxTrip and MaxQuad do the calculations.

Examination of Table 1 of that paper, reproduced below, reveals that the most frequently occurring four character sequence appearing in the Torah occurs 1839 times. This sequence is 'H!H, known as the tetragrammaton, one of the many Divine names.

Table 1 - MaxQuad in the Torah

| ELS SKIP DISTANCE | MAXQUAD |
|---|---|
| 1 | 1839 |
| 2 | 462 |
| 3 | 225 |
| 4 | 192 |
| 5 | 136 |

The number 1839 is interesting for at least two reasons.

Firstly, approximately eight hundred years ago Moses Maimonides painstakingly counted up the number of positive and negative obligations of the Jewish faith which were written into the plain text of the Torah. The total number of obligations is 613, and these comprise the Mitzvot of the Jewish faith. 1839 is, of course, 3 times 613. Both 3 and 613 are prime.

Secondly, we note that the ratio between the rest mass of an electron and the rest mass of a neutron is, to four significant figures, 1839. The three basic constituents of matter in modern physics are the proton, neutron and electron. A skeptic would quite correctly point out that there is no mathematical or statistical reason why 1839 'H!H's appearing in the Torah should be anything other than a complete coincidence.

However, if we take the view that this is not a coincidence we might wonder whether the proton/electron rest mass ratio of 1836 might be indicated in some way or other. A neutron is very slightly heavier than a proton.

### 2.1 The Inverted Letter Nuns

For some reason the Hebrew letter nun is written upside down twice in the text of the Torah. Nobody really knows why the inverted letter nuns are there, and there are a variety of explanations, including one observation that the short section of text between the inverted nuns is a book in its own right [Liebovitz, 2000]. The inverted nuns bracket the two verses of Numbers 10:35-36 and comprise 85 Hebrew characters.

We expect therefore that there might be three 'H!H's in the section of text between the inverted letter nuns. This is not the case - there are only two. In which case we might surmise that one of the 'H!H's must be hidden.

This is in fact what appears to be the case.

If we arrange the characters between the inverted letter nuns in 5 rows of 17, per our *a priori* decryption strategy, we notice that the eighth or middle column contains the characters HH!'A. One of the few ways that these characters can be sensibly re-arranged is such that they spell A'H!H - "One God" in English.

Because there are two H's in 'H!H, there are two ways of re-arranging the rows to achieve the correct spelling of the Divine name prefixed by the letter A. In the following figures the middle column is indicated with an asterisk.

Note that in column 1 of Figure 2, reading bottom to top, rather curiously, we have the word APR'L quite clearly displayed. Coincidentally, figures 1, 2 and 3 were first drawn by the author in April 1999.

Figure 1 - Original text of the Torah

```
                              *
ר מ א י ו נ ר א ה ע ס נ ב י ה י ו   0
א ו צ פ י ו ה ו ה י ה מ ו ק ה ש מ   1
פ מ כ י א נ ש מ ו ס נ י ו כ י ב י   2
י ה ב ו ש מ א י ה ח נ ב ו כ י נ   3
ל א ר ש י פ ל א ת ו ב ר ה ו ה   4
```

Figure 2 - Rows switched - A

```
                              *
ל א ר ש י פ ל א ת ו ב ר ה ו ה   4
י ה ב ו ש מ א י ה ח נ ב ו כ י נ   3
ר מ א י ו נ ר א ה ע ס נ ב י ה י ו   0
פ מ כ י א נ ש מ ו ס נ י ו כ י ב י   2
א ו צ פ י ו ה ו ה י ה מ ו ק ה ש מ   1
```

^ A P R ' L

Figure 3 - Rows switched - B

```
                              *
ל א ר ש י פ ל א ת ו ב ר ה ו ה   4
י ה ב ו ש מ א י ה ח נ ב ו כ י נ   3
א ו צ פ י ו ה ו ה י ה מ ו ק ה ש מ   1
פ מ כ י א נ ש מ ו ס נ י ו כ י ב י   2
ר מ א י ו נ ר א ה ע ס נ ב י ה י ו   0
```

## 2.2 "Receipt" of Key Physical Constants

The factual existence (or "receipt") of these key quantities in an ancient document is the direct reciprocal of similar evidence that humans have transmitted or "sent" in an outbound direction from the Arecibo radio telescope in 1974 [Cornell, 1999] and on the earlier Pioneer 10 & 11 spacecraft in 1972 and 1973 to civilizations unknown [NASA, 1997]. Note that the length of the Arecibo image is a semi prime of 1679 i.e. 23 * 73 bits. Note that the length of the Torah including the inverted letter nuns is both a prime – 304807 and a decimal emirp since 708403 is also prime. We reproduce the Arecibo and Pioneer transmissions and their intended meanings in Appendix One. Both transmissions, though relatively obvious in structure and content to us as senders, would require a significant effort in decoding by life forms that had the even the same biology as us.

Given the presence of such an obvious constant as 1839 (and potentially 1836) in such a precious historical document as the Torah, it is the author's justified belief that we are being signalled to with fundamental, universal, physical constants.

It is furthermore curious that the integer 613 should appear both in the factorization of 1839 and in the number of Mitzvot written within the plain text of the Torah. This could be taken as some form of loose cross check that the content of the plain text of the Torah is correct.

## 3 TWO CONCEPTUALLY RELATED ELS'S IN CLOSE PROXIMITY

The strategy behind WRR's work is to show that conceptually related ELS's appear in close proximity in the Torah. That is to say, they show that sets of ELS word pairs with related meanings (Hammer, Anvil; Chair, Table etc) appear closer together than would be expected by random chance. The concept of a minimal ELS pairing is introduced which deals with the distance between the two words in the pair where each word in the pair appears with a minimal skip. As previously mentioned, statistical methods by Rips, McKay and others are used to approach this controversial matter from the two main opposing perspectives.

Mindful of the time and effort that R'PS (Rips) and QR!LL (Croll) have devoted to the Torah over decades and years respectively, the author was curious to see if these conceptually related words appeared in close proximity in the Torah with a small skip.

### 3.1 R'PS

The minimal ELS for R'PS appears in Leviticus 11:4 with a skip of two. This section of the Torah contains the Laws of Kashrush which are the Jewish dietary laws. Eliyahu Rips is Jewish. R'PS appears again with a slightly larger skip of 5 in Exodus 34:33:

Figure 4 – Minimal ELS for R'PS

```
          v         v         v         v
      ויכל משה מדבר אתמ ויתנ על-פניו מסוה
```

**"Moses finished speaking with them and placed a mask on his face"**

This and subsequent verses talk of the use of a mask which was only removed when Moses was communicating with Hashem (God) or Israel. Although Rips and his colleagues have not yet achieved a widely accepted proof that the Torah is cryptographic material (i.e. its true meaning is concealed as if behind a mask), not even the skeptics would deny that Rips is central to the issue of investigating the matter.

## 3.2 QR!LL

The author became aware of the work of Rips and colleagues some time in 1997 CE and has been applying a variety of quantitative methods to the Torah since April 1998 CE. The minimal ELS for QR!LL appears with a skip of 5 in Exodus 34:2 et seq:

Figure 5 – Minimal ELS for QR'LL

```
       v         v         v         v         v
    והיה נכונ לבקר ועלית בבקר אל-הר
         v         v         v      v
       סיני ונצבת לי שמ על-ראש ההר
               v         v         v         v
    ואיש לא-יעלה עמכ וגמ-איש אל-ירא בכל-ההר

    גמ-הצאנ והבקר אל-ירעו אל-מול ההר ההוא
```

**"Be prepared in the morning; ascend Mount Sinai in the morning and stand by Me there on the mountaintop. No man may ascend with you nor may anyone be seen on the entire mountain. Even the flock and the cattle may not graze facing that mountain".**

The first verses of Exodus Chapter 34 deal with the second set of tablets of stone.

Continuing the QR!LL sequence backwards for three characters we have NBL which would appear to correspond with the last few consonants of the author's Christian name (the author was baptised, confirmed and married within the Church of England Christian tradition). Continuing the QR!LL ELS sequence forwards for five characters yields H$L!$ ("The Trinity").

Regarding the motivation behind the keyword search in this section, I should mention that within my original book [Croll, 1997] I made a suggestion that if the Torah was authored as tradition dictates, and my own book did in fact contain the actual Torah decode sequence, then my name and job title would be encrypted in the Torah in close proximity in the manner demonstrated by WRR.

Thus a near minimal ELS for Rips and a minimal ELS for Croll, two Torah codes proponents, appear in close proximity in Exodus Chapter 34 verses 2 and 33. Continuation of at least one of the ELS's for a few characters forwards and/or backwards confounds these coincidences further. The surface text in the region where each ELS appears has a mysterious resonance with the task at hand.

I thank Prof. Brendan McKay for his comments concerning the number of lameds in the transliteration of my surname as above, however two was my only and prior choice.

## 4 SUMMARY

We have introduced our work investigating the cryptographic properties of the Torah based on an a set of *a priori* principles developed in 1997. We have outlined the main results and their presentation in Jerusalem in June 2000 and in Cambridge in August 2005 together with their subsequent publication on ArXiv in 2007. We have outlined two sets of coincidences, which taken together with our main results, plus the more recent independent scientific work of Prof. Rips, Prof. Haralick and colleagues, suggest that the case for the Torah being a cryptographic object is firm. Furthermore, the idea that characters of the entire Torah might be rearranged into an alternative disposition is not new, there being an extant nomenclature for the process and the individual(s) concerned.

**5 FUTURE WORK**

We show [Croll, 2000a] that algorithmic permutations of the whole Torah produced using Algorithm One are as disordered as one might expect, which reminds us of a quote from Carl Sagan's novel "Contact" [Sagan, 1985] (p50):

> "At other times, like now, when the static was clearly patternless, she would remind herself of Shannon's famous dictum in information theory, that the most efficiently coded message was indistinguishable from noise, unless you had the key to the encoding beforehand."

The obvious problem we have in attempting to algorithmically permute the Torah from its present form into a New Torah (i.e. to decode it to plain) is that of combinatorial complexity.

However, if we are firm in our belief that the Torah is a cryptogram (specifically, a transposition cipher) and it is intended that it be decoded, then in the absence of specific instructions (such as may be given or alluded to in Sefer Yetzirah [Kaplan, 1997]), we have to assume that the cryptographic methodology is natural, and follows relatively easily from the basic principles of mathematics, cryptography and computer science. The most obvious aspect of the methodology being the primality of the length of the Torah.

By way of further example, in our *a priori* decryption principles, we introduce the "Topological Interlock". T2 appears to be topologically interlocked in that, bizarrely as it may seem, the month of rearrangement of T2 appears column wise, as does a hidden tetragrammaton which was also sought. The appearance of further interlocks may guide us further. We note the fun and ingenuity to be enjoyed in the setting and solution of a traditional crossword puzzle

We note that our original strategy for decoding the Torah was to search for lucid permutations of T2, the production parameters for which could be used as parameters to unlock the whole Torah as the dimensionalities of T2 and the whole Torah (T1T2T3) are the same. Unfortunately, the selection of such "lucid" sequences would be entirely subjective and open in terms of the numbers of sequences one might find and have to try, leading to further intractable combinatorial complexity.

Following a decade's contemplation, it turns out that there is an obvious set of permutations of T2 that can be used as a key library. The number of members of this set is relatively small such that the number of potential combinations is finite, countable and potentially tractable. These keys have the additional property of non-commutativity. There is also some evidence that these keys can be sequenced. We hope to outline this key library and their potential application in a following paper.

**APPENDIX ONE**

Figure 6 – The Arecibo Message

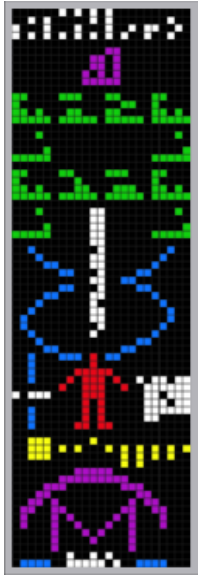

1. The Numbers 1 to 10
2. The Atomic Numbers of the atoms H, C, N, O, P of DNA
3. The formulas for the sugars and bases in the nucleotides of DNA

4. The number of nucleotides in DNA, and a graphic of the double helix structure of DNA

5. A graphic figure of a human, the dimension (physical height) of an average man, and the human population of Earth
6. A graphic of the Solar System, with the Earth offset

7. A graphic of the Arecibo radio telescope and the dimension (the physical diameter) of the transmitting antenna dish

Figure 7 – The Pioneer 10 & 11 Plaques

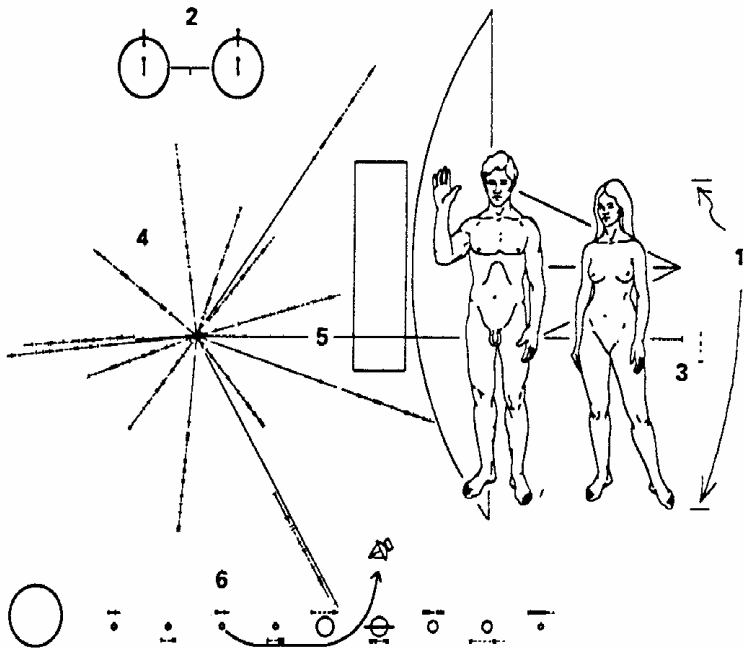

1. Height of man and woman by reference to spacecraft
2. $H_2$ Isospin giving units of length & Time
3. Spacecraft outline
4. Relative position of sun and 14 Quasars
5. Suns relative distance to centre of Galaxy
6. Spacecraft trajectory

**APPENDIX TWO**

Hebrew-English Transliteration used in this paper.

| Hebrew Letter | English Letter | Hebrew Symbol |
|---|---|---|
| Alef | A | א |
| Beit | B | ב |
| Gimmel | G | ג |
| Dalet | D | ד |
| Hey | H | ה |
| Vav | ! | ו |
| Zain | Z | ז |
| Chet | X | ח |
| Tet | T | ט |
| Yud | ' | י |
| Kaf | K | כ |
| Lamed | L | ל |
| Mem | M | מ |
| Nun | N | נ |
| Samech | S | ס |
| Ain | Y | ע |
| Pey | P | פ |
| Tzadi | C | צ |
| Kuf | Q | ק |
| Reish | R | ר |
| Shin | $ | ש |
| Tav | T | ת |